\documentclass{pasj00}

\begin{document}
\SetRunningHead{Yoshida et al.}{Spectropolarimetry of M82}

\title{Spectropolarimetry of the superwind filaments of the starburst 
galaxy M82: kinematics of dust outflow}

\author{%
  Michitoshi \textsc{Yoshida}\altaffilmark{1,2}
  \thanks{Based on data collected with the Subaru Telescope operated
by the National Astronomical Observatory of Japan},
  Koji, S. \textsc{Kawabata}\altaffilmark{1}
  and
  Youichi \textsc{Ohyama}\altaffilmark{3}}
\altaffiltext{1}{Hiroshima Astrophysical Science Center, Hiroshima University,
Higashi-Hiroshima, Hiroshima 739-8526, Japan.}
\email{yoshidam@hiroshima-u.ac.jp}
\altaffiltext{2}{Okayama Astrophysical Observatory, National Astronomical
Observatory of Japan,
Kamogata, Asakuchi, Okayama 719-0232, Japan}
\altaffiltext{3}{Academia Sinica, Institute of Astronomy and Astrophysics,
P.O. Box 23-141, Taipei 10617, Taiwan, R.O.C.}

%


\maketitle

\begin{abstract}
Spectropolarimetry results for the starburst galaxy M82 are presented. 
The optical emission lines of the filaments in the energetic outflow 
(``superwind'') from the nuclear starburst region of M82 are substantially polarized.
The H$\alpha$ polarization degrees and angles measured by our study are 
consistent with previous narrowband imaging polarimetry data.
The polarized emission lines are redshifted with respect to the emission lines 
in the total light and systemic motion of the galaxy.
The emission line intensity ratios [N~{\sc ii}]/H$\alpha$ and [S~{\sc ii}]/H$\alpha$ 
in the polarized light are similar to those of the nuclear star-forming region.
In addition, the electron density $N_{\rm e}$ derived from the
[S~{\sc ii}]$\lambda$6731/$\lambda$6717 line ratio of the polarized light is 
$\sim 600 - 1000$ cm$^{-3}$ at a distance of more than 1 kpc from the nucleus, 
whereas the $N_{\rm e}$ derived from the total light are less than 300 cm$^{-3}$.
These facts strongly suggest that the emission from the nuclear starburst of M82
is scattered by dust grains entrained and transported outward by the superwind.
A simple hollow biconical outflow model shows that the velocity of the 
outflowing dust grains, $v_{\rm d}$, ranges from 100 to 200 km~s$^{-1}$ near the nucleus, 
decreases monotonically with the distance from the nucleus, and reaches $\sim 10$ km~s$^{-1}$ 
at around 1 kpc.
The motion of the dust is substantially slower than that of both ionized gas
($v_{\rm H\alpha} \sim 600$ km~s$^{-1}$) and molecular gas 
($v_{\rm CO} \sim 200$ km~s$^{-1}$) at the same distance from the nucleus of M82.
This indicates that dust grains in the superwind are kinematically decoupled 
from both gas components at large radii.
Since the dust velocity $v_{\rm d}$ is much less than the escape velocity of 
M82 ($v_{\rm esc} \approx 170$ km~s$^{-1}$ at 1.5 kpc from the nucleus), most 
of the dust entrained by the superwind cannot escape to intergalactic 
space, and may fall back into the galaxy disk without any additional 
acceleration mechanisms (such as radiation pressure).
\end{abstract}

\section{Introduction}

Active star formation in starburst galaxies plays one of the most important
roles in galaxy evolution.
In spite of its short duration ($\sim 10^{7-8}$ yrs), a starburst reforms a
substantial amount of the interstellar medium (ISM) of a galaxy into stars.
In addition, a starburst creates an enormous hot outflow from a galaxy via 
the collective effect of supernovae explosions and the stellar winds of massive stars.
This energetic outflow, known as a ``superwind,'' is a ubiquitous phenomenon 
in starburst galaxies (e.g., \cite{heckman03}; \cite{veil05}).
Theoretical studies estimate that the terminal velocity of the hot gas of a
superwind reaches $\sim 10^3$ km s$^{-1}$ (e.g \cite{chev85}),
far exceeding the typical escape velocity of a galaxy.
Metal-rich gas from the starburst region and ambient disk gas are expelled 
from the galaxy disk and pollute the galaxy halo and intergalactic space.
Rapid consumption of the interstellar medium by a superwind will cause star 
formation in a galaxy to cease abruptly,
although part of the expelled gas returns to the disk and induces 
further star formation.
These negative and positive feedbacks greatly affect the chemical evolution 
of a galaxy.

A significant amount of dust is associated with superwinds.
Submillimeter (sub-mm), far-infrared (IR), and mid-IR maps, optical color maps, and polarization studies of several nearby starburst galaxies have exhibited extended dust emission along the 
galaxy minor axis 
(\cite{alton99}; \cite{leeuw09}; \cite{ichi94}; \cite{sca91}; \cite{engel06}; \cite{kaneda10}).
Sub-mm observations of some starburst galaxies suggest that the mass of the 
dust may reach $\sim 10^{6-7}$ M$_{\odot}$ (\cite{alton99}; \cite{leeuw09}).
Although such a large-scale dust outflow has been inferred to play an 
important role in the evolution and metal enrichment of the intergalactic medium 
and halo gas, as well as the evolution of the host galaxy itself, the fate of this 
dust has thus far remained uncertain.
This is because kinematic information on dust outflow is quite 
difficult to obtain, since 
dust produces no sharp emission or absorption lines by which 
its radial velocity can be measured. 

One of the most promising techniques for probing the motion of dust in and 
around superwinds is optical spectropolarimetry.
Outflowing dust grains entrained by a superwind scatter and polarize the 
continuum light and emission lines emanating from the nuclear starburst region.
In other words, the dust grains act as ``moving mirrors'' for nuclear light.
Hence, the velocity measured by the polarized emission lines of a superwind 
must reflect the motion of the dust with respect to the nucleus. 
Motivated by this idea, we carried out deep optical spectropolarimetry of the 
prototypical starburst galaxy M82 to reveal the dust kinematics of 
its superwind.

M82 is distinguished by its very bright, kpc-scale superwind 
(\cite{nakai87}; \cite{sea01}; \cite{matsu05}; \cite{bierao08}; \cite{bland88}; 
\cite{shop98}; \cite{ohyama02}; \cite{mut07}; \cite{breg95}; \cite{tsuru07}; 
\cite{rana08}; \cite{strick07}).
The superwind of M82 is accompanied by large extraplanar dust filaments, as well
as hot ionized gas (\cite{ichi94}; \cite{alton99}; \cite{thuma00}).
\citet{alton99} found a huge dust envelope extending along the minor axis of M82 
using the 850-$\mu$m sub-mm observation.
Recently, \citet{leeuw09} detected a much fainter dust emission as far as 1.5 kpc 
from the galaxy disk.
They found that the sub-mm morphology has a north--south asymmetry, which is 
consistent with the H$\alpha$ and X-ray morphologies.
Mid- and far-IR maps recently obtained using IR space telescopes reveal a 
complex structure in the kpc-scale filaments of polycyclic aromatic 
hydrocarbon (PAH) dust extending along the minor axis of M82 (\cite{engel06}; 
\cite{kaneda10}; \cite{roussel10}).
The highly polarized nature of the optical continuum and H$\alpha$ emission of the 
outer region of M82 also indicates the presence of a vast quantity of dust in the superwind (\cite{schmidt76}; \cite{bing76}; \cite{vis72}; \cite{sca91}).
The polarization degree of the H$\alpha$ emission reaches 30\%\ in some areas, 
and the polarization angle is almost perpendicular to 
the radial direction drawn from the nucleus of the galaxy \citep{sca91}.
Therefore, M82 is an ideal object for studying dust kinematics in a starburst 
superwind via optical spectropolarimetry.

In this paper, we present the results of deep optical spectropolarimetric observations 
of the superwind of M82.
This is the first attempt to ascertain the spatial structure
of the dust motion in a starburst superwind via spectropolarimetry.
We adopted 3.89 Mpc as the distance to M82 \citep{sakai99}, which yields a 
linear scale of 18.9 pc~arcsec$^{-1}$ for the galaxy.

\section{Observations}

Spectropolarimetric observations of M82 were made with FOCAS \citep{kashik02}, 
attached to the Cassegrain focus of the Subaru Telescope \citep{kaifu00}, on 
December 22 and 23, 2003.
The observations were carried out using the spectropolarimetric mode of FOCAS \citep
{kawabata03}.
We used a slit mask with eight 0\arcsec.6 (width) $\times$ 
20\arcsec.6 (length) slitlets at 23\arcsec.7 intervals,
and a VPH grism with 665 grooves~mm$^{-1}$ and a center wavelength of 6500 \AA.
The resultant spectral resolving power was $\lambda / \Delta\lambda \approx 1700$, 
determined by the combination of the slit and the grating. The separation direction of the
beam splitter was perpendicular to the direction of the slit length, and 
spectra of both ordinary and extraordinary rays were obtained simultaneously.

\begin{figure}
  \begin{center}
    \FigureFile(90mm,80mm){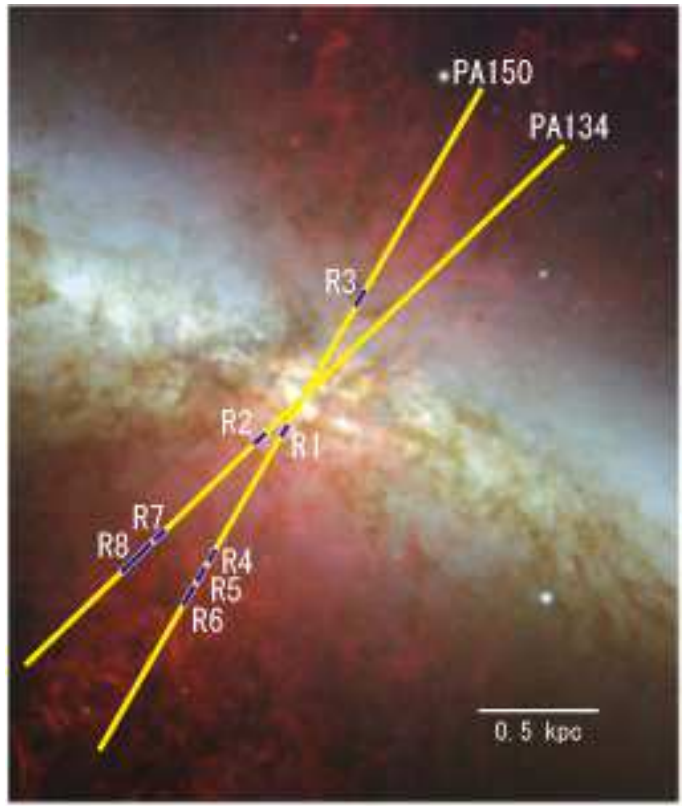}
  \end{center}
  \caption{The slit positions overlaid on a false color (blue:$V$, green:$R$, and 
red:H$\alpha$) image of M82.
Regions discussed in the text are labeled R1--R8.}
           \label{slit}
\end{figure}

\begin{figure*}
  \begin{center}
    \FigureFile(160mm,80mm){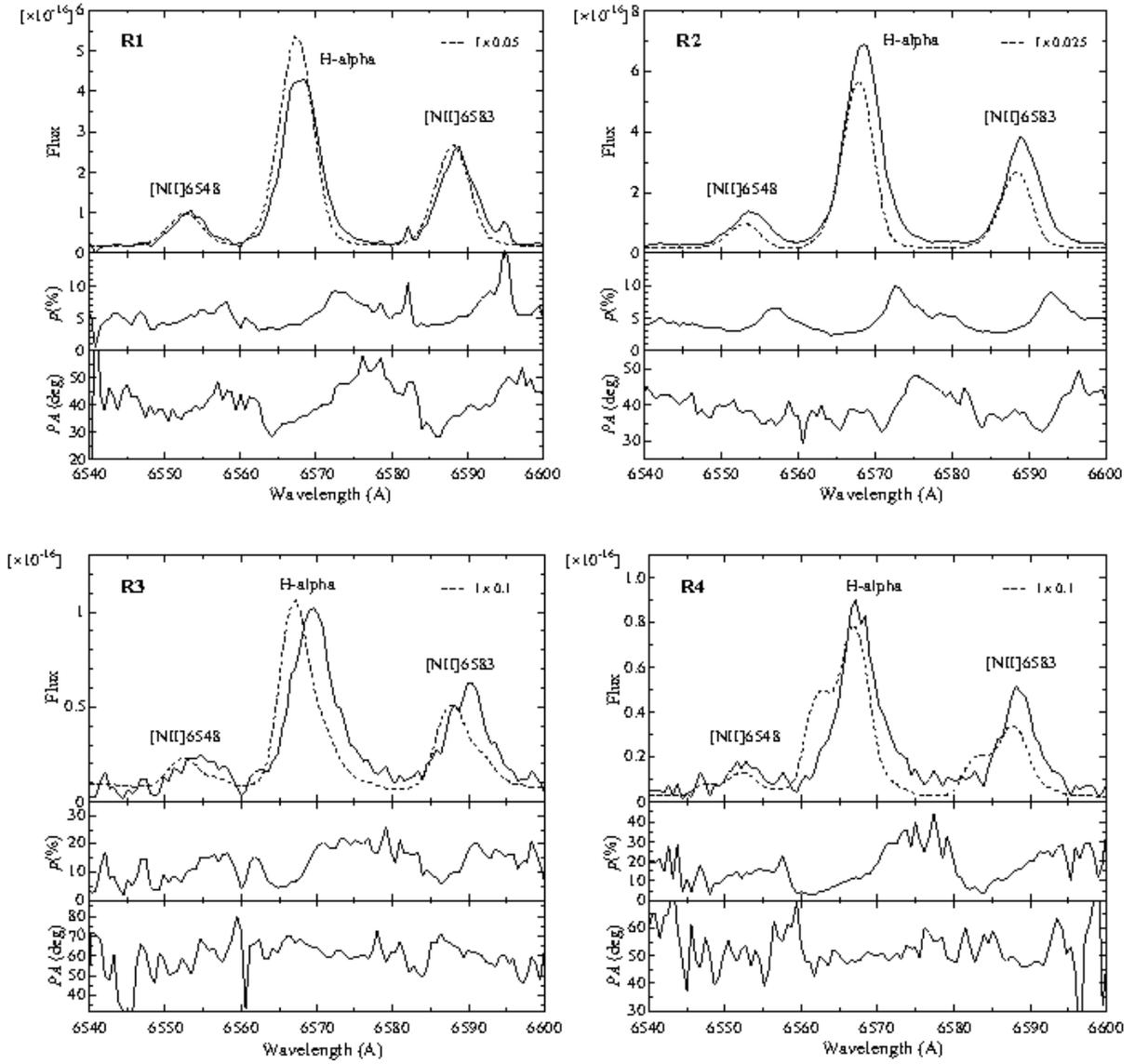}
  \end{center}
  \caption{Spectra of total light and polarized light around the
H$\alpha$+[N~{\sc ii}] wavelength region for R1--R4.
The polarized spectra are represented by the solid line in the upper panels of each figure.
The total light spectra are scaled by a factor shown in the upper-right corner 
of each figure and are represented by the dashed line.
The polarization degree and polarization position angle are shown in the 
middle and bottom panels of each figure, respectively.}
           \label{hasp-1}
\end{figure*}

\begin{figure*}
  \begin{center}
    \FigureFile(160mm,80mm){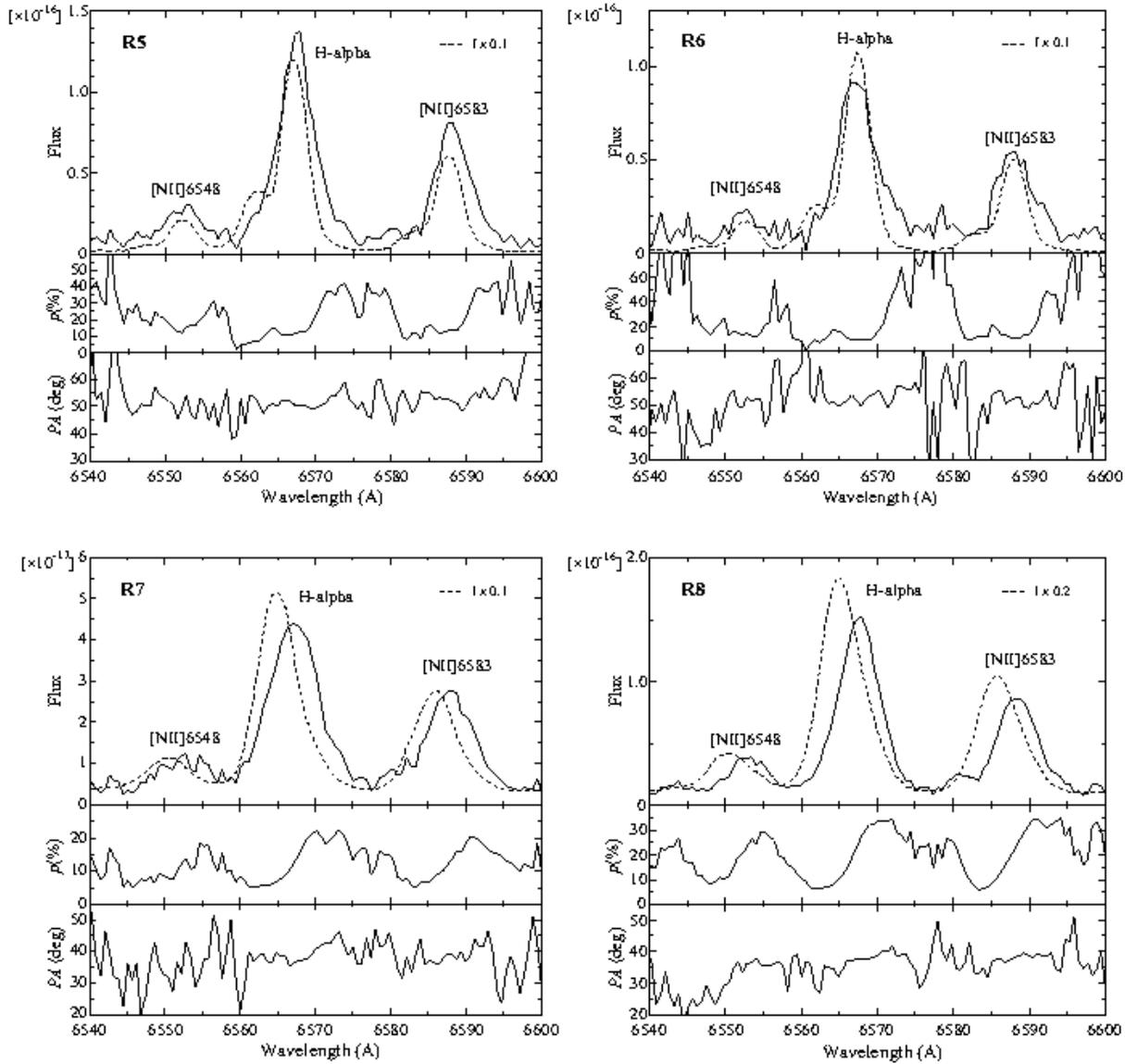}
  \end{center}
  \caption{Spectra around the H$\alpha$+[N~{\sc ii}] wavelength region for R5--R8.
The layout is the same as in Figure 2.}
           \label{hasp-2}
\end{figure*}

\begin{figure*}
  \begin{center}
    \FigureFile(160mm,80mm){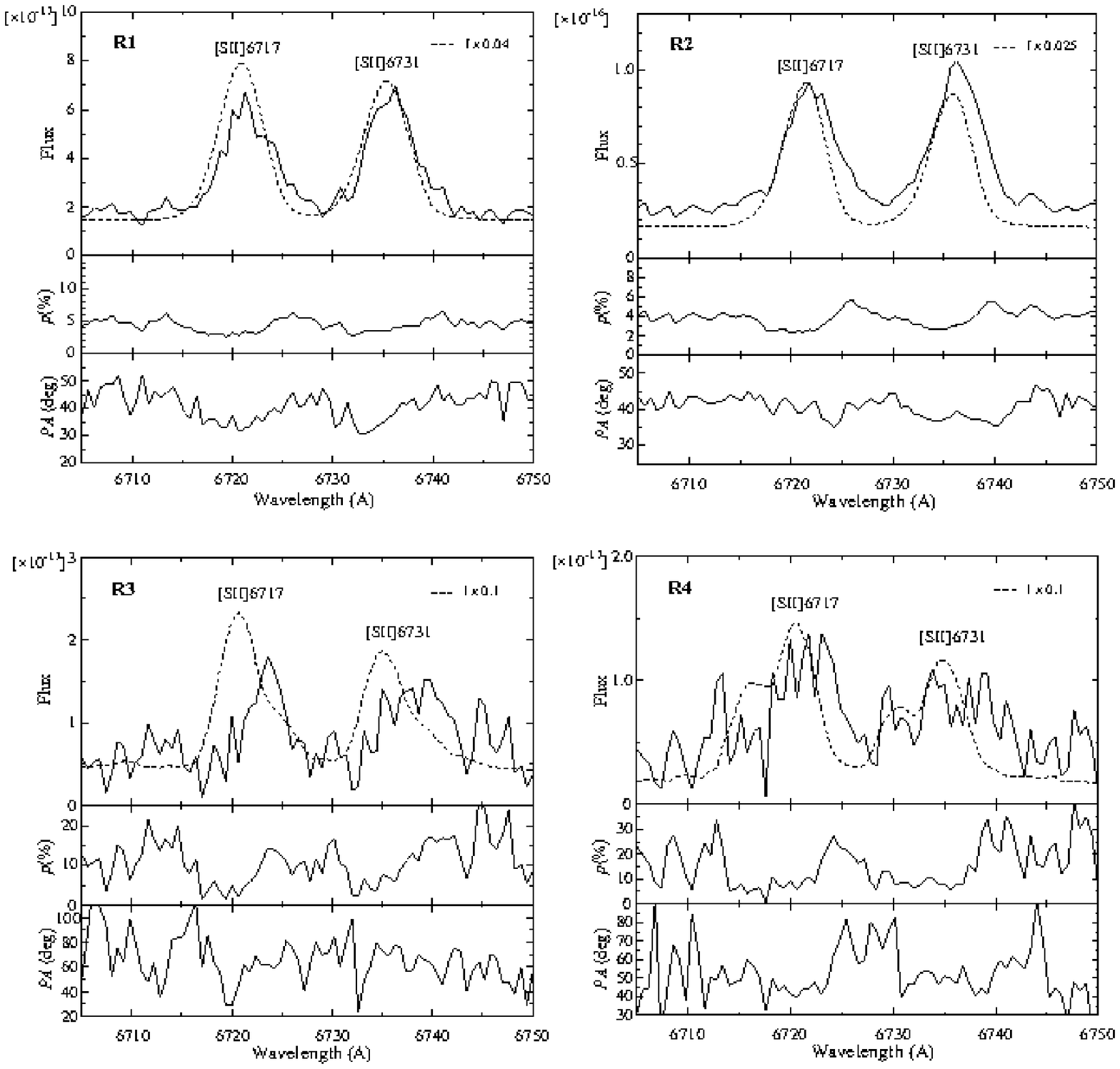}
  \end{center}
  \caption{Spectra around the [S~{\sc ii}] wavelength region for R1--R4.
The layout is the same as in Figure 2.}
           \label{s2sp-1}
\end{figure*}

\begin{figure*}
  \begin{center}
    \FigureFile(160mm,80mm){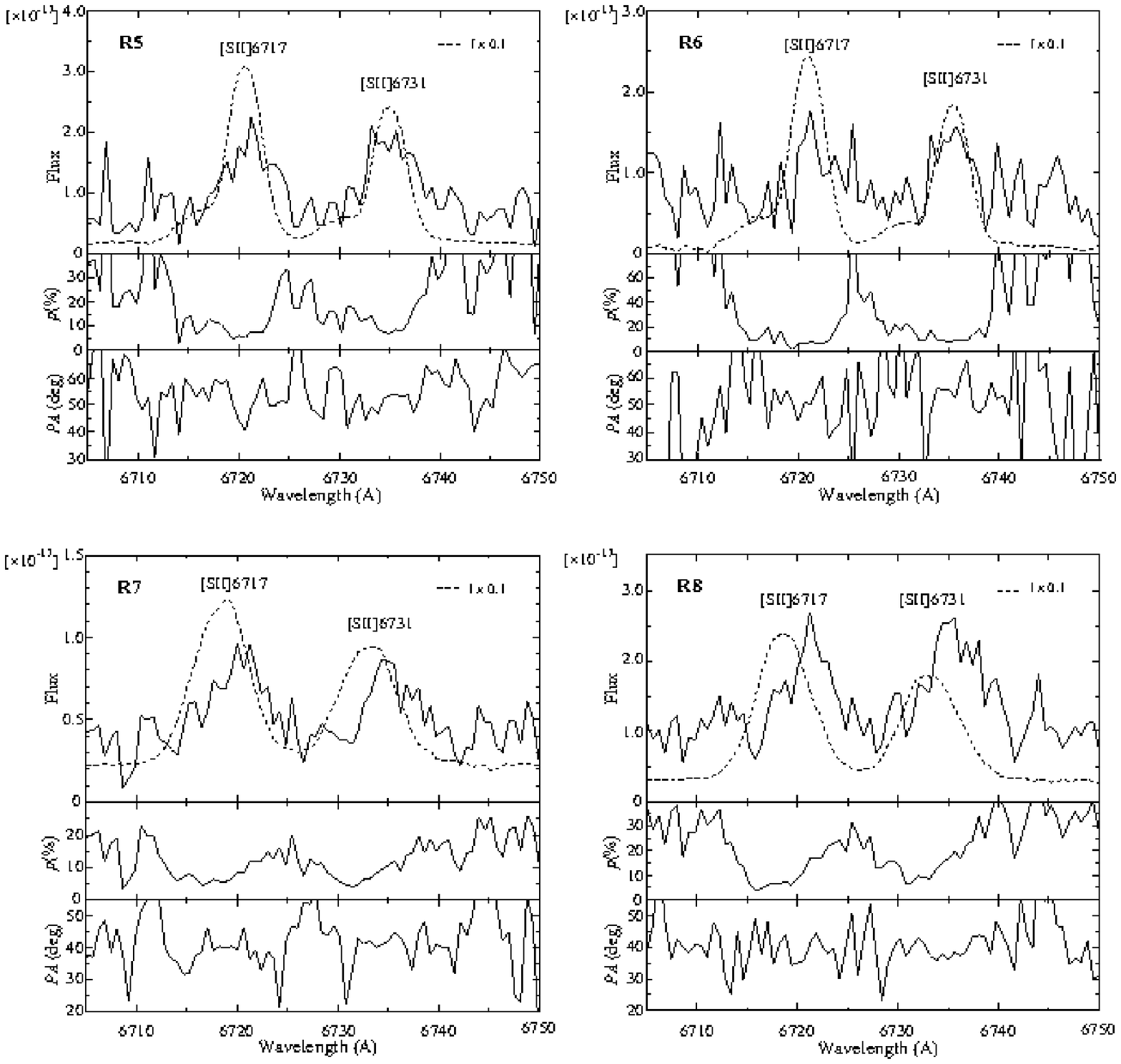}
  \end{center}
  \caption{Spectra around the [S~{\sc ii}] wavelength region for R5--R8.
The layout is the same as in Figure 2.}
           \label{s2sp-2}
\end{figure*}

The PA of the slit was set at 150$^\circ$ and 134$^\circ$.
We placed the slit center at the position of the 2.2 $\mu$m nucleus \citep{tele91}.
A unit data set consists of exposures
taken at four different position angles (PAs; 0$^\circ$, 
22.5$^\circ$, 45$^\circ$, 67.5$^\circ$) of the half-wave plate.
We took one data set with an exposure time of 600 s per exposure for PA 150$^\circ$, and
four data sets with an exposure time of 720 s per exposure for PA 134$^\circ$.
Thus, the total exposure times were 2400 s for PA 150$^\circ$ and 11520 s for PA 
134$^\circ$.
Figure \ref{slit} shows the positions of the slits overlaid on an 
image taken with FOCAS \citep{ohyama02}.
The regions in which the spectra were extracted and summed are also shown in 
Figure \ref{slit}.

We observed both an unpolarized star and a highly polarized star to calibrate the
polarization.
We also observed a spectrophotometric standard star.
The first night of the observing run, December 22, was clear and photometric.
The second night was not photometric, with some small passing clouds. 
The seeing was approximately 0\arcsec.6 the first night and 0\arcsec.7--
0\arcsec.8 the second night.

\section{Data Reduction}

Standard CCD data reduction was applied to the raw CCD frames using IRAF.
The bias level was subtracted from all frames.
Since the dark current was negligible ($<$ 1 ADU), dark subtraction was not 
performed.
Flat fielding was carried out using the dome flat data averaged over the four PAs 
of the half-wave plate.

We extracted one-dimensional spectra of ordinary and extraordinary rays
in eight regions (R1--R8) of the two-dimensional spectral images.
The locations of R1--R8 are shown in Figure \ref{slit}. The distances from the 
nucleus and the areas of the regions are summarized in Table 1.  
We estimated the sky background components from the two-dimensional spectra 
in the outer region (140 -- 145\arcsec\ from the nucleus), where no significant 
H$\alpha$ flux was recorded.
The polarization parameters were calculated from eight spectra
(ordinary and extraordinary at the four wave-plate PAs) employing
the method described in \S 6.1.2 of \citet{tin96}.
Our observations of unpolarized stars indicated that 
instrumental polarization was negligible ($\lesssim 0.1$\%).
Moreover, our measurements of flat-field lamps through fully polarizing
filters showed that the depolarization factor was also negligible
($\lesssim 0.05$). Therefore, we made no correction for 
instrumental polarization and depolarization.
The zero point of the position angle on the sky was determined 
from the observation of the highly polarized star.

We fitted Gaussian profiles to the emission line spectra for both the total 
light and the polarized light.
First we applied Gaussian fitting to the H$\alpha$+[N~{\sc ii}] spectra, 
assuming that H$\alpha$, [N~{\sc ii}]$\lambda$6548, and [N~{\sc ii}]$\lambda$6583
have the same radial velocities and FWHMs.
We fixed the emission line intensity ratio [N~{\sc ii}]$\lambda$6583/$\lambda$6548
at 3 for this procedure.
The H$\alpha$+[N~{\sc ii}] emission lines of the total light in R3, R7, and R8 
have a remarkable red-wing component, while those in R4, R5, and R6 exhibit a
double-peak profile.
We fitted double Gaussian profiles to the H$\alpha$+[N~{\sc ii}] lines of these 
regions to decompose their asymmetric profiles. 
The radial velocities and FWHMs of all components were assumed to be
the same for the decomposition.
We then applied Gaussian fitting to the [S~{\sc ii}]$\lambda\lambda$6717,6731 
lines, assuming their radial velocities and FWHMs to be the same as those of the 
H$\alpha$+[N~{\sc ii}] lines.
Some asymmetric features were also found in the polarized emission line
profiles of R3--R7, but the S/N ratios were not high enough to obtain a
reliable decomposition.
Hence we applied a single Gaussian fitting to all the polarized emission lines,
and then corrected for instrumental broadening by applying the following simple
equation: ${\rm FWHM} = \sqrt{{\rm FWHM}_{\rm fit}^2 - {\rm FWHM}_{\rm inst}^2}$,
where FWHM$_{\rm fit}$ and FWHM$_{\rm inst}$ are the FWHM obtained from the fitting and
that of the instrumental profile, respectively.
We determined that FWHM$_{\rm inst} = 140$ km~s$^{-1}$ by fitting the sky emission 
lines near H$\alpha$.
The physical parameters derived by the above procedure are listed in Tables 1 and 2.

\section{Properties of the polarized emission lines}

\begin{longtable}{lcccccccccc}
  \caption{Physical Parameters of the H$\alpha$-emitting Gas in the M82 Superwind}\label{tab:T1}
  \hline
  ID & PA$^{\ast}$ & D$^{\dagger}$ & Area & 
  $f_{\rm H\alpha}$$^{\ddagger}$ &
  $v$ & FWHM & 
  $f_{\rm [N~{\sc II}]}$$^{\ddagger}$ &
  $f_{\rm [S~{\sc II}]6717}$$^{\ddagger}$ &
  $f_{\rm [S~{\sc II}]6731}$$^{\ddagger}$ &
  $N_{\rm e, t}$ \\
   & (deg) & (pc) & (pc$\times$pc) & & (km~s$^{-1}$) & (km~s$^{-1}$) &  &  &  & (cm$^{-3}$) \\ 
\endfirsthead
  \hline
\endfoot
  \hline
  \multicolumn{10}{l}{\footnotesize $^{\ast}$ The position angle of the slit.}
\\
  \multicolumn{10}{l}{\footnotesize $^{\dagger}$ The projected distance from 
the nucleus.} \\
  \multicolumn{10}{l}{\footnotesize $^{\ddagger}$ The emission line flux in 
units of $10^{-15}$ erg~s$^{-1}$~cm$^{-2}$.} \\
\endlastfoot
  \hline
    R1 & 150 & 140 & 11$\times$28 & 97 & $215\pm10$ & $189\pm22$ & 46 & 15 &
  13 & $370\pm10$ \\
    R2 & 134 & 245 & 11$\times$32 & 190 & $234\pm10$ & $161\pm26$ & 85 & 26 &
  25 & $450\pm10$ \\
    R3 & 150 & 630 & 11$\times$38 & 7.5 & $192\pm11$ & $150\pm28$ & 3.3 & 1.4 &
 1.1 & $120\pm10$ \\
       &     &     &              & 2.3 & $385\pm12$ & $224\pm19$ & 1.4 & 0.4 &
 0.4 & $370\pm10$ \\
    R4 & 150 & 735 & 11$\times$28 & 6.7 & $195\pm11$ & $187\pm23$ & 2.9 & 1.1 &
 0.85 & $110\pm10$ \\
       &     &     &              & 2.6 & $-32\pm14$ & $75\pm63$ & 0.83 & 0.41 &
 0.31 & $120\pm10$ \\
    R5 & 150 & 805 & 11$\times$32 & 8.7 & $205\pm11$ & $121\pm35$ & 4.4 & 2.1 &
 1.6 & $120\pm10$ \\
       &     &     &              & 2.4 & $-43\pm13$ & $95\pm46$ & 0.67 & 0.31 &
 0.26 & $260\pm10$ \\
    R6 & 150 & 910 & 11$\times$43 & 7.7 & $215\pm11$ & $116\pm37$ & 3.6 & 1.7 &
 1.2 & $70\pm10$ \\
       &     &     &              & 1.7 & $-40\pm13$ & $107\pm41$ & 0.57 & 0.24 &
 0.21 & $320\pm10$ \\
    R7 & 134 & 840 & 11$\times$29 & 3.0 & $83\pm10$ & $178\pm23$ & 1.2 & 0.75 &
 0.50 & $<50$ \\
       &     &     &              & 2.5 & $167\pm11$ & $325\pm12$ & 1.7 & 0.41 &
 0.37 & $400\pm10$ \\
    R8 & 134 & 980 & 11$\times$77 & 3.0 & $75\pm14$ & $130\pm33$ & 1.3 & 0.98 &
 0.61 & $<50$ \\
       &     &     &              & 7.0 & $155\pm11$ & $295\pm14$ & 4.3 & 1.4 &
 1.2 & $260\pm10$ \\
\end{longtable}

Figures \ref{hasp-1} and \ref{hasp-2} show the total fluxes, polarized 
fluxes, polarization degrees, and polarization angles of 
the H$\alpha$+[N~{\sc ii}] regions of the eight selected regions along the slits (R1--R8).
Figures \ref{s2sp-1} and \ref{s2sp-2} show the same data for the [S~{\sc ii}] 
region of the spectra.
We summarize the physical parameters derived from the total light spectra in Table 1.
Since double Gaussian fittings were applied to the emission line profiles of R3--R8, 
the fitted parameters for each component are listed separately for these regions 
in Table 1.  
The emission line fluxes of the polarized H$\alpha$, [N~{\sc ii}]$\lambda$6581
and [S~{\sc ii}]$\lambda\lambda$6717,6731, the velocities and FWHMs of the
polarized H$\alpha$, and the electron densities $N_{\rm e}$ derived from the 
polarized [S~{\sc ii}] line intensity ratios are given in Table 2.
We assumed an electron temperature of $10^4$ K in calculating $N_{\rm e}$.
The radial velocities are corrected to heliocentric values.

The intensity-weighted polarization degrees of the emission lines range from 
5\%\ to 15\%\ (see Table 2).
The polarization degree begins to decrease on the blue side of the emission 
lines, reaches its minimum near the center of the lines, increases again on the red 
side of the lines, and attains a peak at the red wings of the lines.
Except for R1 and R2, the polarization angles are almost constant 
($\approx 40^{\circ}$ -- $60^{\circ}$) throughout the H$\alpha$+[N~{\sc ii}] spectral 
region, and no difference exists between the polarization angles of the 
continuum and emission lines.
In the spectra of R1 and R2, the polarization angles increase toward the red 
part of the emission lines (upper panels of Figure \ref{hasp-1}).
The polarization angles are $\approx 35^\circ$ in the continuum and in the blue 
part of the emission lines, while the angles reach  
$\approx 50^\circ$ -- $60^\circ$ at the red wings of the lines.
Except for the fine structures seen in R1 and R2, the polarization vectors are generally
 perpendicular to radial lines drawn from the M82 nucleus.

\begin{figure}
  \begin{center}
    \FigureFile(95mm,50mm){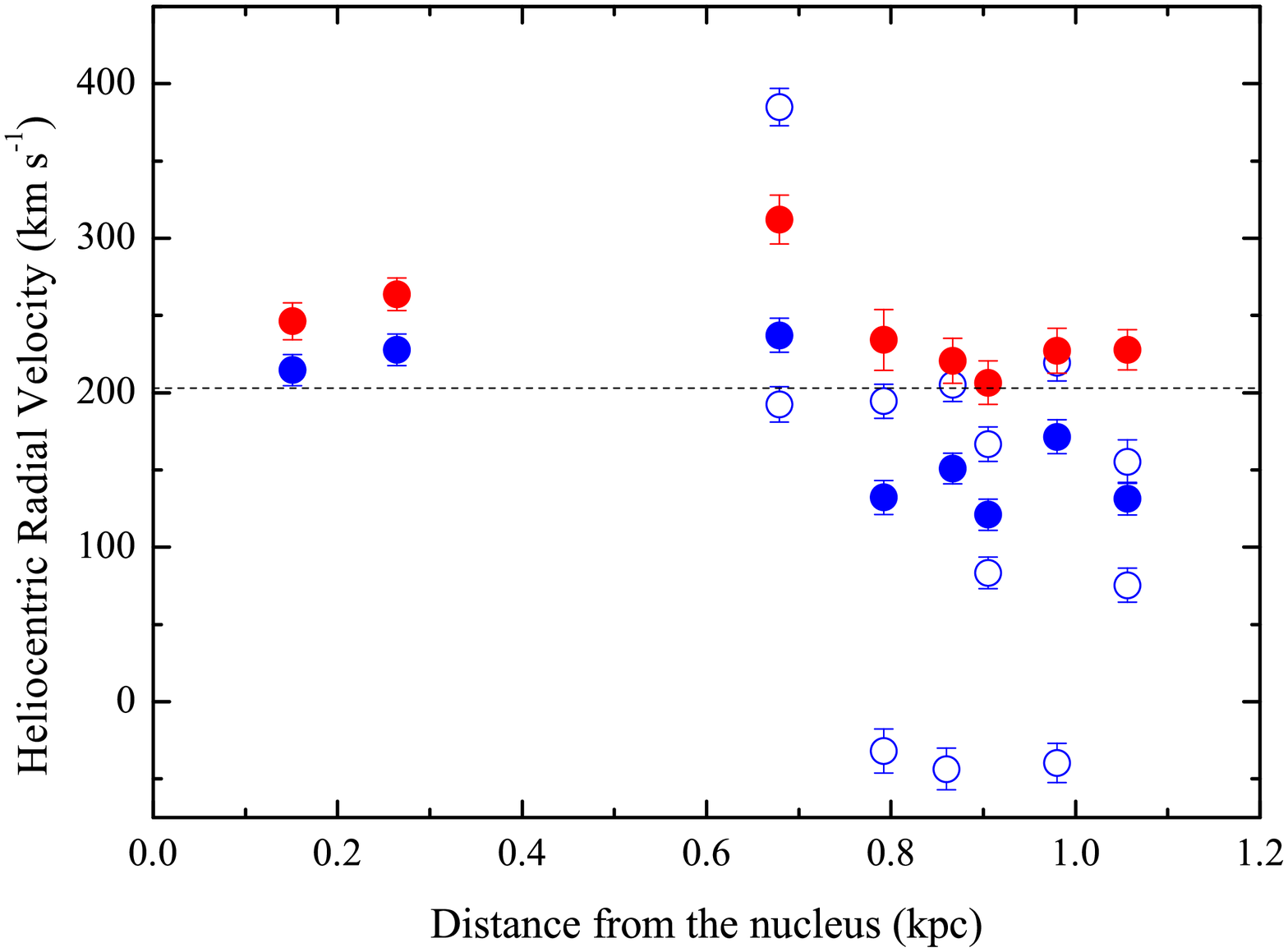}
  \end{center}
  \caption{Position--velocity diagram of the H$\alpha$ emission of the superwind 
of M82. 
The filled red and blue circles indicate the heliocentric radial velocities of the polarized 
H$\alpha$ emission and the intensity-averaged velocities of the H$\alpha$ emission for
the total light, respectively.
The profiles of the total light emission lines in the outer regions (R3--R8) are
fitted with double Gaussian functions, and the velocities of the split
components are then plotted as open blue circles.
The dotted line represents the systemic velocity of M82.
 }
           \label{pvel}
\end{figure}

\begin{longtable}{*{10}{c}}
  \caption{Physical Parameters of the Gas Measured in Polarized Emission Lines of the M82 Superwind}\label{tab:T2}
  \hline
  ID & 
  P.Deg.$^{\ast}$ &
  P.PA$^{\dagger}$ & 
  $f_{\rm H\alpha, p}$$^{\ddagger}$ & 
  $v_{\rm p}$ & FWHM & 
  $f_{\rm [N~{\sc II}]6583, p}$$^{\ddagger}$ & 
  $f_{\rm [S~{\sc II}]6717, p}$$^{\ddagger}$ & 
  $f_{\rm [S~{\sc II}]6731, p}$$^{\ddagger}$ & 
  $N_{\rm e,p}$ \\ 
   & (\%) & (deg)  & &  (km~s$^{-1}$) & (km~s$^{-1}$) & & & & (cm$^{-3}$) \\
  \hline
  \endhead
  \hline
  \endfoot
  \hline
  \multicolumn{10}{l}{\footnotesize $^{\ast}$ The intensity-weighted 
mean polarization degree of the H$\alpha$ emission.} \\
  \multicolumn{10}{l}{\footnotesize $^{\dagger}$ The intensity-weighted
mean position angle of the polarization vector of the H$\alpha$ emission.} \\
  \multicolumn{10}{l}{\footnotesize $^{\ddagger}$ The polarized emission line
flux in units of $10^{-16}$ erg~s$^{-1}$~cm$^{-2}$.} \\
  \endlastfoot
  \hline
    R1 & 4.5 & 35 & 44 & $252\pm11$ & $219\pm19$ & 24 & 4.2 & 4.8 & $890\pm10$ \\
    R2 & 3.4 & 37 & 65 & $247\pm11$ & $203\pm21$ & 33 & 6.7 & 7.6 & $910\pm10$ \\
    R3 & 12.7 & 65 & 12 & $317\pm14$ & $279\pm15$ & 6.4 & 0.70 & 0.67 & $520\pm50$ \\
    R4 & 10.5 & 49 & 9.8 & $240\pm15$ & $268\pm16$ & 4.6 & 0.51 & 0.40 & $130\pm50$ \\
    R5 & 11.4 & 51 & 12 & $226\pm13$ & $221\pm19$ & 6.9 & 1.2 & 1.2 & $740\pm40$ \\
    R6 & 9.8 & 54 & 8.6 & $227\pm15$ & $221\pm19$ & 4.1 & 0.44 & 0.47 & $750\pm130$ \\
    R7 & 10.4 & 38 & 5.6 & $213\pm14$ & $312\pm14$ & 3.0 & 0.50 & 0.44 & $310\pm40$ \\
    R8 & 15.4 & 36 & 15 & $234\pm13$ & $240\pm18$ & 8.1 & 1.3 & 1.5 & $940\pm20$ \\
\end{longtable}

Figure \ref{pvel} shows the radial velocities of the total light and 
polarized light versus the distance from the nucleus.
Clearly the polarized emission line spectra are redshifted with respect to the 
total spectra (see also Figures \ref{hasp-1} - \ref{s2sp-2}).
Also, the polarized H$\alpha$ lines have radial velocities higher than the 
systemic velocity of M82 ($v_{\rm sys} = 203$ km~s$^{-1}$; \cite{shop98};
\cite{gotz90}).
The velocity excess, $v_{\rm p} - v_{\rm sys}$, ranges from $\sim 
10$ km~s$^{-1}$ to $\sim 100$ km~s$^{-1}$.
In the regions close to the nucleus (R1 and R2) $v_{\rm p} - v_{\rm sys} 
\approx 50$ km~s$^{-1}$, increasing to $\approx 100$ km~s$^{-1}$ in R3.
Note that R3 is in the northwest part of the superwind of M82.
In the southeast regions (R4--R8), whose distances from the nucleus are 
0.7 -- 1 kpc, the polarized H$\alpha$ lines have almost constant velocities, 
and $v_{\rm p} - v_{\rm sys} \sim 10$ km~s$^{-1}$.
In contrast, the H$\alpha$ lines of the total light are blueshifted 
relative to $v_{\rm sys}$ in R4--R8, which is consistent with previous studies 
(\cite{shop98}; \cite{greve04}).

\begin{figure}
  \begin{center}
    \FigureFile(95mm,50mm){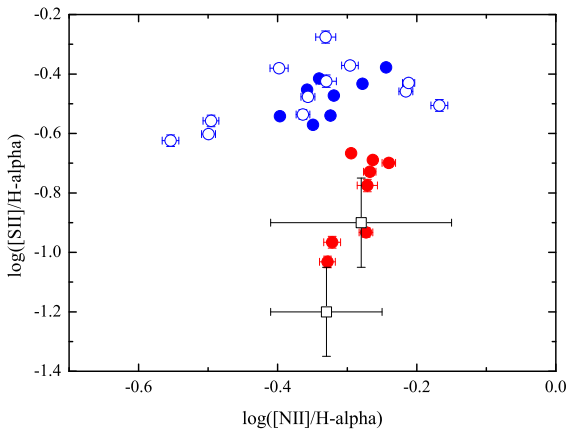}
  \end{center}
  \caption{The emission line intensity ratios of the polarized light and the total light
[S~{\sc ii}](6717+6731)/H$\alpha$ vs. [N~{\sc ii}]6584/H$\alpha$.
The filled red and blue circles indicate the data for the polarized light and 
the total light, respectively.
The open blue circles represent the data for the split components of the total
light.
The open squares show the data in an area of $\approx 50 \times 100$ pc around the 
nucleus of M82, measured by integral field spectroscopy \citep{west09}.
The upper and lower squares represent the data for the narrow component and the 
broad component, respectively (``C1'' and ``C2'' in \citet{west09}).
The error bars associated with the squares show the distribution of the data 
over the measured region.}
           \label{excitation}
\end{figure}

The emission line intensity ratios [N~{\sc ii}]/H$\alpha$ and 
[S~{\sc ii}]($\lambda$6717+$\lambda$6731)/H$\alpha$ for the polarized light are 
similar to those of the nuclear region.
Figure \ref{excitation} shows a plot of the ratios for the total light and the 
polarized light, together with the data taken around the nucleus  
\citep{west09}.
The data for the polarized light are distributed in and around the range of the 
nuclear data.
The data for the total light, however, are distributed well outside 
the nuclear data (Figure \ref{excitation}).

\begin{figure}
  \begin{center}
    \FigureFile(95mm,50mm){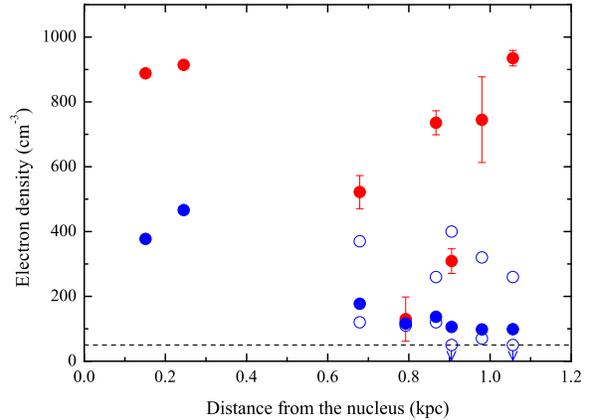}
  \end{center}
  \caption{Electron density derived from the [S~{\sc ii}] emission lines of M82.
The filled red and blue circles indicate the data for the polarized light and 
total light, respectively.
The open blue circles represent the data for the split components of the
total light.
The dashed line marks the lower limit of electron density that can be
derived from the intensity ratios of the [S~{\sc ii}] emission lines.}
           \label{eden}
\end{figure}

The emission line intensity ratios [S~{\sc ii}]$\lambda$6731/
[S~{\sc ii}]$\lambda$6717 of the polarized spectra 
are higher than those of the total light in all regions except R4 
(Figures \ref{s2sp-1} and \ref{s2sp-2}).
Figure \ref{eden} shows the $N_{\rm e}$ derived from the [S~{\sc ii}] line ratios.
The electron densities $N_{\rm e, t}$ obtained from the [S~{\sc ii}] line ratios of 
the total light are $\approx 400$ cm$^{-3}$ at $\sim 0.2$ kpc from the nucleus 
and decrease to $< 300$ cm$^{-3}$ at $>$0.5 kpc (Figure \ref{eden}).
In contrast, the electron densities $N_{\rm e, p}$ measured in the polarized spectra
 are almost constant ($\approx$ 500 -- 1000 cm$^{-3}$), except in the two low-density regions 
(R4 and R7). 
This indicates that scattered light originates from a high-density region.
The electron density of the nucleus of M82 is $\approx 1000$ cm$^{-3}$ \citep{west09}.
The $N_{\rm e, p}$ we derived are slightly smaller than, but almost consistent with, 
the nuclear $N_{\rm e}$.

The foregoing polarized light characteristics (polarization angles, 
emission line ratios, and $N_{\rm e, p}$ measured in the polarized light)
 are consistent 
with the idea that the polarized light is nuclear light scattered
by the dust in the superwind of M82.
In addition, we discovered that the radial velocities of the polarized emission
lines are systematically greater than the velocity $v_{\rm sys}$ of the galaxy.
The dust grains that scatter the nuclear light can be inferred to move
 outward from the nucleus.
Accordingly, our research strongly suggests that dust grains entrained by the superwind
are transported outward from the disk by the wind and scatter the nuclear light.

\section{Outflow velocity of the dust wind}

\begin{figure}
  \begin{center}
    \FigureFile(85mm,60mm){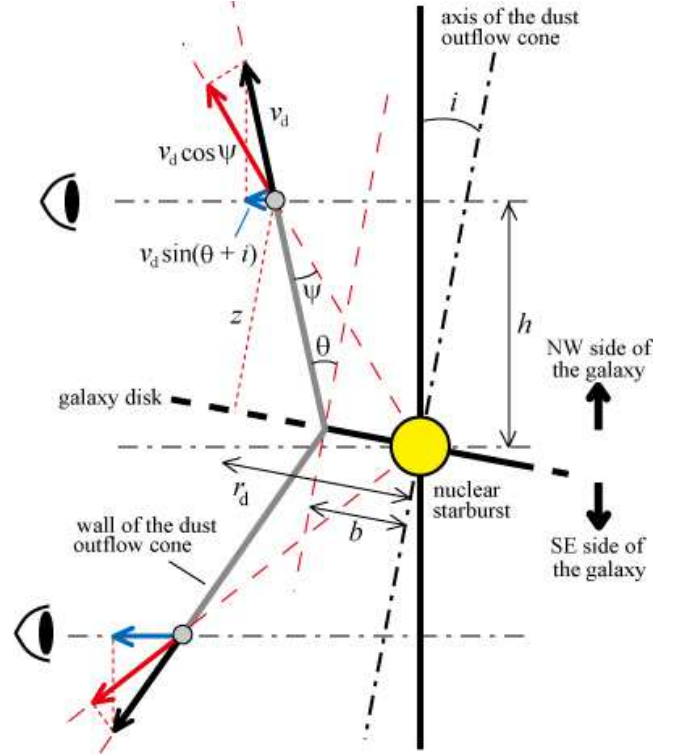}
  \end{center}
  \caption{A schematic diagram of the M82 dust wind. 
The dust grains in the galaxy disk are assumed to be entrained by the 
superwind expelled from a circumnuclear circular 
region whose radius in the disk is $b$.
The entrained dust flows along the walls of a cone whose opening angle is $2 \times 
\theta$.
The angle of inclination of the axis of the dust flow cone is $i$.
The dust grains in the wind reflect the nuclear emission, acting as mirrors moving at
velocity $v_{\rm d} {\rm cos} \psi$ with respect to the nucleus, where $v_{\rm d}$ is 
the outflow velocity of the dust.
}
           \label{model}
\end{figure}

We attempt to derive the true outflow velocity of the dust grains in the superwind of M82 
using the polarized spectra.
Assuming that the radial velocities of the polarized emission lines reflect the 
motion of the dust grains with respect to the nucleus, we construct a 
hollow biconical outflow model for the dust outflow of M82 (Figure \ref{model}).
Since forward scattering is much more efficient for scattering by ordinary 
interstellar dust grains \citep{hulst57}, we ignored the backscattering component 
from the far sides of the outflow cones.

The idea of the model is as follows.
Dust grains are flowing outward from the galaxy disk along the walls of the 
hollow cones, with outflow velocity $v_{\rm d}$.
The radius of the base of the outflow cone is $b$ kpc.
The dust grains act as moving mirrors, scattering the light from the nucleus.
The opening angle of the outflow is $2 \times \theta$, and the angle of inclination 
of the axis of the outflow cone is $i$ (Figure \ref{model}).
Although $i$ is generally different from the angle of inclination of the 
galaxy disk (i.e., the axis of the cone may be tilted with respect to the rotational axis of 
the galaxy), the difference between the two values should be small, as suggested 
by the ionized gas wind morphology (\cite{shop98}; \cite{ohyama02}; \cite
{greve04}).
Hence we consider $i$ to be effectively the same as the inclination of the galaxy.
In the following calculation, the sign of $i$ is negative on the northwestern
side of the galaxy and positive on the southeastern side.
When the dust grains are observed at a projected height $h$ above the 
nucleus, the distance $z$ between the observed dust grains and the galaxy 
disk (``true height'') is expressed as follows:
\begin{equation}
z = \frac{h + b\; {\rm tan} (i)\; {\rm cos} (i)}{1 - {\rm tan} (\theta)\; {\rm 
tan} (i)} .
\end{equation}
The distance $r_{\rm d}$ between the position of the dust grains projected onto the 
galaxy disk and the nucleus (i.e., the radial distance within the galaxy disk)
 is then $r_{\rm d} = z \; {\rm tan} ( \theta ) + b$.
Using these values, we can express the angle $\psi$ between the outflow velocity
$v_{\rm d}$ and the line connecting the observed flow region and the nucleus as 
\begin{equation}
\psi = \pi / 2 - \theta - {\rm tan}^{-1} \left( \frac{z}{r_{\rm d}} \right).
\end{equation}
The dust grains are receding from the nucleus with a velocity
$v_{\rm d}\;{\rm cos}(\psi)$ (red arrows in Figure \ref{model}), 
and the scattered emission lines are redshifted with this velocity.
In contrast, the velocity component of the grains along the line of sight
is $v_{\rm d}\;{\rm sin}(\theta + i)$ (blue arrows in Figure \ref{model}),
and hence the scattered emission lines, are blueshifted with this velocity,
relative to the systemic motion of the galaxy.
As a result, the observed radial velocity $v_{\rm p}$ of the scattered emission
lines will be $v_{\rm p} = v_{\rm d}\;{\rm cos} (\psi) - \{ v_{\rm d}\;{\rm sin}(\theta + i) - v_{\rm sys}\} $. 
In this simple model, we can thus derive the dust outflow velocity $v_{\rm d}$ as 
follows:
\begin{equation}
v_{\rm d} = \frac{v_{\rm p} - v_{\rm sys}}{{\rm cos} (\psi) - {\rm sin} 
( \theta + i )} ,
\end{equation}
where $v_{\rm p}$ and $v_{\rm sys}$ ($= 203$ km~s$^{-1}$; \cite{gotz90}) are the radial velocity of the polarized emission lines and the systemic velocity of M82, respectively.

To calculate $v_{\rm d}$, we make the following assumptions about 
the parameters, $b$, $\theta$, and $i$.
The dust can be reasonably expected to be entrained from the galaxy disk outside the 
main hot outflow.
The radius of the base of the dust outflow is then assumed to be the same 
as the radius of the central starburst,
so that $b \approx 0.3$ kpc \citep{greve04}.
The opening angle $2\theta$ and the angle of inclination $i$ of the dust outflow are
 hardly determined only by our observations.
We examine two cases for $\theta$: a narrow-angle case and a wide-angle case.
In the narrow-angle case, we assume that the dust grains move in the same 
direction as the ionized gas.
In this case, we adopt the value $2\theta = 25^{\circ}$, which was derived by 
\citet{greve04}.  
If the dust flow is associated with molecular gas, $\theta$ may be much wider 
than the ionized gas flow (the wide-angle case).
When the dust grains are associated with the molecular outflow 
identified by \citet{walter02}, $2\theta \approx 50^\circ$.
\citet{ohyama02} suggested a wider opening angle for the dust outflow in or near 
the galaxy disk ($2\theta > 90^\circ$) using an optical extinction map (see also
\cite{ichi94}).
However, the dust scattering the nuclear light is very far away from 
the disk according to the imaging polarimetry maps \citep{sca91}.
Hence the wider-angle dust flows near the disk plane would not contribute to
the polarized light we detected here.
Thus, we adopt the value $2 \theta = 50^{\circ}$ in the wide-angle case.
Finally, we assume that $i = 15^{\circ}$, which was derived by \citet{greve04}, 
for both cases.

\begin{figure}
  \begin{center}
    \FigureFile(100mm,80mm){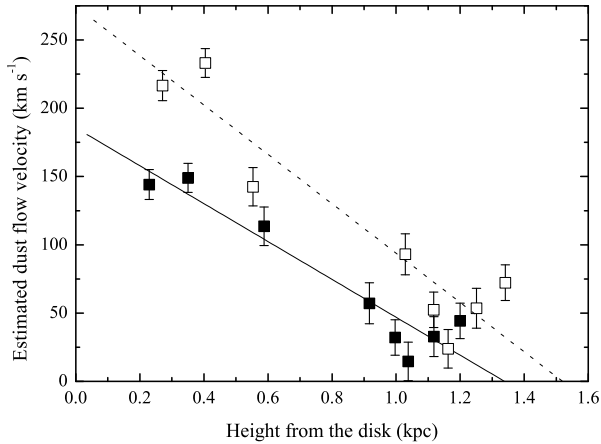}
  \end{center}
  \caption{Estimated dust outflow velocity $v_{\rm d}$ of M82. 
The filled squares and open squares represent the data derived for the 
narrow-angle case and the wide-angle case, respectively.
The horizontal axis denotes the height measured from the galaxy disk ($z$ in 
Figure 9).
The solid and dotted lines represent best-fit curves of
constant deceleration models (see text) for the narrow-angle case 
and the wide-angle case, respectively.}
           \label{dvel1}
\end{figure}

We find that the estimated dust outflow velocities based on the above 
simple model are much less than the escape velocity of M82.  
In Figure \ref{dvel1}, the dust outflow velocities $v_{\rm d}$ of all regions are plotted against the 
height $z$ measured from the galaxy disk for the above two cases.
The dust outflow velocity $v_{\rm d}$ decreases almost monotonically with $z$
 and reaches a minimum value of $\sim$ 20 -- 30 km~s$^{-1}$ at 
approximately 1--1.2 kpc in both cases.
\citet{martin98} estimated the escape velocity of M82 using the H~{\sc i} velocity 
field and found that $v_{\rm esc} \approx 170$ km~s$^{-1}$ at 1.5 kpc from the 
nucleus \citep{martin98}.
The present estimate for $v_{\rm d}$ is much less than this value.
Some indication of an upturn in $v_{\rm d}$ for $z \gtrsim 1.2$ kpc
is observed (see Figure 10),
but whether this trend reflects a global acceleration of the 
dust grains at high altitudes above the disk is not clear, owing to the lack of data beyond 
$z \sim 1.5$ kpc.

We fit a constant deceleration model to the dust velocity field
in the M82 superwind and estimate the dust recycling timescale.
Assuming that the dust grains are entrained by the superwind at an initial 
velocity $v_{\rm d, 0}$ and are decelerated at a constant rate 
$d_{\rm r}$, we fit linear functions $v_{\rm d} (z) = v_{\rm d,0} + d_{\rm r} z$ for the 
two cases and find ($v_{\rm d,0}$ km~s$^{-1}$, $d_{\rm r}$ 
km~s$^{-1}$~kpc$^{-1}$) $=$ ($186$, $-139$) for the narrow-angle 
case and ($268$, $-176$) for the wide-angle case.
We can then calculate the distance $z(t)$ (in kpc) attained by the dust grains 
at time $t$ (in Myr) from 
$z(t) = v_{\rm d, 0}/d_{\rm r} \cdot\{1 - {\rm exp} (- 1.1\times10^{-3}\cdot d_{\rm r}\; t ) \}$,
since the grains are entrained from the galaxy disk.
In this simple model, the dust grains reach a height of 1 kpc above the 
disk in $\sim 10^7$ yr in both cases.
If the deceleration rate $d_{\rm r}$ is constant, $v_{\rm d}$ would equal zero at 
$\sim 1.5$ kpc above the disk.
The free-fall timescale from 1.5 kpc above the disk of M82 to the disk plane 
is calculated as 2 -- 3$\times 10^7$ yr, under the assumption that a mass of
$\sim 10^{10}$ M$_\odot$ is confined within 2 kpc from the M82 nucleus \citep{sofue92}. 
Hence, if the deceleration of the dust flow is monotonic, the dust 
expelled by the superwind may return to the disk in several times 10$^7$ yr.
Note that this timescale is comparable to that of starburst activity.

\section{Radial extension of the dust outflow of M82}

Our simple dust outflow model suggests that dust grains cannot go
beyond 1.5 kpc above the disk of M82.
However, a body of evidence indicates that dust actually distributes 
over a few kpc around M82.
The imaging polarimetry observations of \citet{sca91} showed that the 
H$\alpha$ emission is highly polarized even at the edge of the ionized gas flow 
of M82 \citep{sca91}.
At the tip of the H$\alpha$ emission region, $\sim 3$ kpc from the disk, the 
polarization degree reaches 30\%, which means that a significant amount of 
dust exists at that altitude.
Sub-mm observations have detected dust emission extending more than a few kpc 
from the disk (\cite{leeuw09}).
An 8-$\mu$m image taken with the $Spitzer$ infrared satellite revealed a 
complex filamentary dust structure extending 3 kpc from the disk 
\citep{engel06}.
Recently, very extended extraplanar cold dust emission in M82 
was discovered via mid- and far-IR observations (\cite{kaneda10}; \cite{roussel10}).
The UV emission extending to 6 kpc along the minor axis of M82 has also
been attributed to dust scattering \citep{hoopes05}.

How dust grains are transported to such a high altitude remains a mystery.
If the dust grains originate from the galaxy disk, additional mechanisms for 
their outward acceleration (such as radiation pressure from the starburst or 
further mechanical acceleration by the wind) are needed to explain such an
extended dust distribution.
\citet{fer91} calculated the effect of radiation pressure on the motion of 
interstellar dust grains.
They determined that dust grains are expelled to intergalactic space 
at several hundred km~s$^{-1}$ by the radiation pressure of stellar light in 
our Galaxy and the star-forming galaxy NGC 3198.
Their calculation predicted that the velocity of the dust would rapidly 
increase to $\sim 500$ km~s$^{-1}$ within 10 Myr from the onset of the flow and 
then decrease to $\sim$100 -- 200 km~s$^{-1}$ in 100 Myr.
The dust grains are transported outward to an order of 100 kpc in 100 Myr.

The results of \authorcite{fer91}'s calculation \citep{fer91} clearly contradict
our observational results for M82.
Our observations suggest that $v_{\rm d}$ decreases with 
time after the onset of the flow (see Section 5).
Note that 
radiation pressure is highly dependent on the dust grain size $a$.
Radiation pressure is
proportional to $a^2$, whereas gravitational force is proportional to $a^3$.
Although the grain size distribution and spatial distribution of the 
extraplanar dust of M82 are not known,
dust components traced by mid-IR observations (\cite{engel06}; 
\cite{bierao08}; \cite{kaneda10}) should consist primarily of small PAH dust particles.
These small dust particles may be transported very far away from the disk by radiation 
pressure and form a highly protracted structure.
Dust particles involved in optical scattering, however, are much larger 
and heavier than PAH dust; sub-micron-sized grains are mostly 
responsible for optical scattering ($2\pi a / \lambda \sim 1$;
e.g., \cite{hulst57}).
Thus, the discrepancy between mid-IR dust extension and optical
scattering dust kinematics is possibly reconciled by the size difference between
the grains responsible for each emission.

Note, however, that the H$\alpha$ emission is scattered even at the tip of the bright
H$\alpha$ filaments \citep{sca91}, indicating that optical scattering dust is still present 
at this distance.
At least two possible explanations exist for this finding:
1) The origin of the outer dust is the same as that of the inner dust, and
the grains are accelerated by some mechanism(s) at $\geq 1$ kpc. 
2) The origin of the outer dust is different from that of the inner dust.
Some indication of dust acceleration is observed at around 1.2 kpc from the
disk, since a velocity upturn appears in the velocity field we obtained (see the previous
section).
This may indicate that the dust grains are accelerated outward beyond 1 kpc.
However, the velocity upturn is not especially significant, since the lack of 
data points makes it impossible to be certain whether the dust is really accelerated
beyond 1 kpc. 
However, the distant dust may be associated with the
H~{\sc i} stream traversing the disk of M82 \citep{yun93}.
This stream is part of a giant H~{\sc i} cloud complex extending across the M81/M82 group,
and possibly formed by a galaxy--galaxy interaction between M82 and M81
(\cite{apple81}; \cite{chyno08}).
Recently, Roussel et al. (\yearcite{roussel10}) detected a protracted cold dust
emission around M82 via far-IR (250 $\mu$m, 350 $\mu$m and 500 $\mu$m) observation
with the $Herschel$ satellite.
The cold dust extends to 9 kpc from the center.
By comparing their far-IR maps with other wavelength data, they found that
the outer part of the cold dust is associated with the H~{\sc i} stream, while
the inner part is dominated by the warm ($\sim 30 - 50$ K) component
associated with the superwind \citep{roussel10}.
In this case, optical scattering dust far away from the disk is provided by
this stream, and the dust velocity field in the outer part of the superwind
may reflect H~{\sc i} stream kinematics.
To distinguish between the two possibilities listed above, one must 
trace the motion of the dust toward a higher altitude above the disk.

\section{Comparison with other outflow materials}

We compare the kinematics of the dust and other outflowing materials, 
including ionized gas, molecular gas, and neutral atomic gas, in the superwind of M82.

The kinematical dissociation between the dust and the warm ionized gas (optical 
emission line gas) of M82 is very clear.
The outflow velocity of the ionized gas increases almost linearly with the distance
from the disk at a rate of 300 -- 400 km~s$^{-1}$~kpc$^{-1}$ (\cite{greve04}).
In contrast, the dust outflow decelerates at a rate of $-150$ km~s$^{-1}$~kpc$^{-1}$
(see Section 5). 
The final outflow velocity of the ionized gas in M82 reaches $\sim$500 -- 600 km~s$^{-1}$
at $\sim 1$ kpc from the disk (\cite{shop98}; \cite{greve04}),
which is much higher than the dust velocity we derived.

The velocity field of the molecular gas in the M82 superwind is also inconsistent
with the kinematics of the dust. 
Several CO observations have detected a molecular gas outflow with velocity $\sim 
200$ km s$^{-1}$ in M82
(\cite{sea01}; \cite{walter02}).
This value is almost the same as the dust velocity near the nucleus ($z < 0.4$ 
kpc) for both the narrow- and wide-angle cases of our dust outflow model.
Hence dust and CO gas may be associated with each other near the nucleus.
The spatial coincidence between CO emission and the 450-$\mu$m emission also 
supports a spatial association of dust and molecular gas in the central 
region \citep{hughes94}.
However, the CO outflow velocities ($\sim 200$ km s$^{-1}$) were measured at
$\sim 1$ kpc from the 
nucleus, where the dust velocity decreases to less than 100 km~s$^{-1}$ (Figure \ref{dvel1}).
Thus, we can conclude that molecular gas is also kinematically decoupled from 
the dust at high altitudes in the superwind of M82.

Finally, we compare dust motion with neutral atomic gas motion.
\citet{heckman00} discovered that Na~{\sc i} absorption lines are systematically blueshifted 
with respect to the systemic velocities of starburst galaxies (see also 
\authorcite{martin02} \yearcite{martin02}, \yearcite{martin05};
\authorcite{rupke02} \yearcite{rupke02}, \yearcite{rupke05}; 
\cite{shw04}).
They interpreted this phenomenon as a neutral gas outflow from the starburst region.
Using a correlation between the optical absorption coefficient $A_V$ and the
equivalent width of the Na~{\sc i} absorption line, they inferred that dust is 
associated with neutral atomic gas, which is represented by Na~{\sc i} absorption.
The outflow velocity derived from the blueshift of the Na~{\sc i} lines of M82 is 
$\approx 80$ km~s$^{-1}$.
This value is significantly smaller than $v_{\rm d}$ near the nucleus, which
may imply that neutral atomic gas is also kinematically decoupled
from the dust near the nucleus.
However, we should note that \citet{heckman00} measured the line-of-sight velocity of
Na~{\sc i} only in the nuclear region.
If the outflow were spherically symmetric, the Na~{\sc i} velocity they derived 
would equal the actual outflow velocity.
However, various observations indicate that the superwind of M82 has a 
bipolar nature.
Hence the Na~{\sc i} velocity represents not the extraplanar outflow velocity, 
but the expansion velocity of the circumnuclear gas in the galactic plane.
In other words, currently available data do not allow us to directly compare the
kinematics of dust and neutral atomic gas.
Deep spectroscopy of Na~{\sc i} absorption lines at 
large radii in M82 would help us to study the detailed outflow kinematics of 
neutral atomic gas and to compare it with dust kinematics.

\section{Conclusion}

We conducted optical spectropolarimetry observations of the starburst superwind 
of the prototypical starburst galaxy M82 to reveal the motion of the dust 
entrained by the superwind.
The H$\alpha$ polarization degrees ($\sim$5 -- 15\%) and angles 
measured by our study are consistent with previous narrowband imaging 
polarimetry maps.
The polarized emission lines are redshifted relative to the systemic motion of 
the galaxy.
The emission line intensity ratios [N~{\sc ii}]/H$\alpha$ and 
[S~{\sc ii}]/H$\alpha$ in the polarized light are similar to those of 
the nuclear star-forming region. 
The electron densities $N_{\rm e}$ derived from the polarized [S~{\sc ii}] line 
ratio are much higher than those derived from the total light.
These facts strongly suggest that the emission from a nuclear starburst is 
scattered by dust grains entrained and then transported outward by the 
starburst superwind.
We derived the outflow velocity of the dust grains, 
$v_{\rm d}$, using a simple hollow biconical outflow model.
The outflow velocity $v_{\rm d}$ is on the order of a few hundred km~s$^{-1}$ near
the nucleus and decreases monotonically with the distance from the nucleus.
The dust motion revealed by this study is substantially slower than the motion 
of the other components of the superwind (ionized gas and molecular gas).
The outflow velocity of the dust is also much less than the escape velocity of 
M82.
In the absence of any additional effective acceleration mechanisms (such as radiation 
pressure), the dust expelled by the superwind would fall back into the galaxy 
disk within several times $10^7$ yr.

\bigskip

We are grateful to the Subaru Telescope staff for their kind
assistance with the observations. 
We also thank the anonymous referee for his/her helpful comments.
Part of this study was carried out using the facilities of the Astronomical
Data Center, National Astronomical Observatory of Japan.
This research made use of NASA's
Astrophysics Data System Abstract Service. This work was financially
supported in part by the Japan Society for the Promotion of Science
(Grant-in-Aid for Scientific Research No. 18340055) and the Ministry
of Education, Culture, Sports, Science \& Technology, Japan (Grant-in-Aid
for Scientific Research on Priority Areas No. 19047003).



\begin{thebibliography}{}
\bibitem[Alton et al.(1999)]{alton99} 
  Alton, P. B., Davies, J. I. \& Bianchi, S. 1999, \aap, 343, 51
\bibitem[Appleton, Davies \& Stephenson(1981)]{apple81}
  Appleton, P. N., Davies, R. D. \& Stephenson, R. J. 1981, \mnras, 195, 327
\bibitem[Bier$\tilde{\rm a}$o et al.(2008)]{bierao08}
  Bier$\tilde{\rm a}$o, P. et al. 2008, \apj, 676, 304
\bibitem[Bingham et al.(1976)]{bing76}
  Bingham, R. G. et al. 1976, \nat, 259, 463
\bibitem[Bland \& Tully(1988)]{bland88}
  Bland, J. \& Tully, R. B. 1988, \nat, 334, 43
\bibitem[Bregman et al.(1995)]{breg95}
  Bregman, J. N., Schulman, E. \& Tomisaka, K. 1995, \apj, 439, 155
\bibitem[Chevalier \& Clegg(1985)]{chev85}
  Chevalier, R. A. \& Clegg, A. W. 1985, \nat, 317, 44
\bibitem[Chynoweth et al.(2008)]{chyno08}
  Chynoweth, K. M. et al. 2008, \aj, 135, 1983
\bibitem[Engelbracht et al.(2006)]{engel06}
  Engelbracht, C. W. et al. 2006, \apjl, 642, L127
\bibitem[Ferrara et al.(1991)]{fer91}
  Ferrara, A., Ferrini, F., Franco, J. \& Barsella, B. 1991, \apj, 381, 137
\bibitem[G$\ddot{\rm o}$tz et al.(1990)]{gotz90}
  G$\ddot{\rm o}$tz, M., McKeith, C. D., Downes, D. \& Greve, A. 1990, \aap, 
240, 52
\bibitem[Greve(2004)]{greve04}
  Greve, A. 2004, \aap, 416, 67
\bibitem[Heckman et al.(2000)]{heckman00}
  Heckman, T. M. et al. 2000, \apjs, 129, 493
\bibitem[Heckman(2003)]{heckman03}
  Heckman, T. M. 2003, RevMexAA, 17, 47
\bibitem[Hoopes(2005)]{hoopes05}
  Hoopes, C. G. et al. 2005, \apjl, 619, L99
\bibitem[Hughes et al.(1994)]{hughes94}
  Hughes, D. H., Gear, W. K \& Robson, E. I. 1994, \mnras, 270, 641
\bibitem[Ichikawa et al.(1994)]{ichi94}
  Ichikawa, T. et al. 1994, \apj, 433, 645
\bibitem[Kaifu et al.(2000)]{kaifu00}
  Kaifu, N., et al. 2000, \pasj, 52, 1
\bibitem[Kaneda et al.(2010)]{kaneda10}
  Kaneda et al. 2010, \aap, 514, A14
\bibitem[Kashikawa et al.(2002)]{kashik02}
  Kashikawa, N., et al. 2002, \pasj, 54, 819
\bibitem[Kawabata et al.(2003)]{kawabata03}
  Kawabata, K. et al. 2003, SPIE, 4841, 1219 
\bibitem[Leeuw \& Robson(2009)]{leeuw09}
  Leeuw, L. L. \& Robson, E. I. 2009, \aj, 137, 517
\bibitem[Martin(1998)]{martin98}
  Martin, C. L. 1998, \apj, 506, 222
\bibitem[Martin et al.(2002)]{martin02}
  Martin, C. L., Kobulnicky, H. A. \& Heckman, T. M. 2002, \apj, 574, 663
\bibitem[Martin(2005)]{martin05}
  Martin, C. L. 2005, \apj, 621, 227
\bibitem[Matsushita et al.(2005)]{matsu05}
  Matsushita, S. et al. 2005, \apj, 618, 712
\bibitem[Mutchler et al.(2007)]{mut07}
  Mutchler, M. et al. 2007, \pasp, 119, 1
\bibitem[Nakai et al.(1987)]{nakai87}
  Nakai, N. et al. 1987, \pasj, 39, 685
\bibitem[Ohyama et al.(2002)]{ohyama02}
  Ohyama, Y. et al. 2002, \pasj, 54, 891
\bibitem[Ranalli et al.(2008)]{rana08}
  Ranalli, P., Comastri, A., Origlia, L. \& Maiolino, R. 2008, \mnras, 386, 1464
\bibitem[Roussel et al.(2010)]{roussel10}
  Roussel, H. et al. 2010, \aap, 518, L66
\bibitem[Rupke et al.(2002)]{rupke02}
  Rupke, D. S., Veilleux, S. \& Sanders, D. B. 2002, \apj, 570, 588
\bibitem[Rupke et al.(2005)]{rupke05}
  Rupke, D. S., Veilleux, S. \& Sanders, D. B. 2005, \apjs, 160, 115
\bibitem[Sakai \& Madore(1999)]{sakai99}
  Sakai, S. \& Madore, B. F. 1999, \apj, 526, 599
\bibitem[Scarrott et al.(1991)]{sca91}
  Scarrott, S. M., Eaton, N. \& Axon, D. J. 1991, \mnras, 252, 12
\bibitem[Schmidt et al. (1976)]{schmidt76}
  Schmidt, G. D., Angel, J. R. P. \& Comwell, R. H. 1976, \apj, 206, 888
\bibitem[Schwartz \& Martin(2004)]{shw04}
  Schwartz, C. M. \& Martin, C. L. 2004, \apj, 610, 201
\bibitem[Seaquist \& Clark(2001)]{sea01}
  Seaquist, E. R. \& Clark, J. 2001, \apj, 552, 133
\bibitem[Seaquist et al.(2006)]{sea06}
  Seaquist, E. R., Lee, S. W. \& Moriarty-Schieven, G. H. 2006, \apj, 638, 148
\bibitem[Shopbell \& Bland-Hawthorn(1998)]{shop98}
  Shopbell, P. L. \& Bland-Hawthorn, J. 1998, \apj, 493, 129
\bibitem[Sofue, Y.(1992)]{sofue92}
  Sofue, Y. et al. 1992, \apj, 395, 126
\bibitem[Strickland \& Stevens(2000)]{strick00}
  Strickland, D. K. \& Stevens, I. R. 2000, \mnras, 314, 511
\bibitem[Strickland \& Heckman(2007)]{strick07}
  Strickland, D. K. \& Heckman, T. M. 2007, \apj, 658, 258
\bibitem[Taylor, Walter \& Yun(2001)]{taylor01}
  Taylor, C. L., Walter, F. \& Yun, M. S. 2001, \apjl, 562, L43
\bibitem[Telesco et al.(1991)]{tele91}
  Telesco, C. M. et al. 1991, \apj, 369, 135
\bibitem[Thuma et al.(2000)]{thuma00}
  Thuma et al. 2000, \aap, 358, 65
\bibitem[Tinbergen(1996)]{tin96} Tinbergen, J. 1996,
  Astronomical Polarimetry (New York: Cambridge Univ. Press)
\bibitem[Tsuru et al.(2007)]{tsuru07}
  Tsuru, T. G. et al. 2007, \pasj, 59, 269
\bibitem[van de Hulst(1957)]{hulst57}
  van de Hulst, H. C. 1957, Light Scattering by Small Particles (New York: J. Wiley \& Sons)
\bibitem[Veilleux et al.(2005)]{veil05}
  Veilleux, S., Cecil, G. \& Bland-Hawthorn, J. 2005, \araa, 43, 769
\bibitem[Visvanathan \& Sandage(1972)]{vis72}
  Visvanathan, N. \& Sandage, A. R. 1972, \apj, 176, 57
\bibitem[Walter et al.(2002)]{walter02}
  Walter, F., Wei\ss, A. \& Scoville, N. Z. 2002, \apjl, 580, L21
\bibitem[Wei\ss et al.(2005)]{weiss05}
  Wei\ss, A, Walter, F. \& Scoville, N. Z. 2005, \aap, 458, 533
\bibitem[Westmoquette et al.(2009)]{west09}
  Westmoquette, W. S. et al. 2009, \apj, 706, 1571
\bibitem[Yun et al.(1993)]{yun93}
  Yun, M. S. et al. 1993, \apjl, 411, L17


\end{thebibliography}
\end{document}